\documentstyle[preprint,aps,prd,latexsym,amssymb]{revtex}

\begin{document}
\tightenlines
\draft

\preprint{Preprint UTF - 409}
\title{Quantum implementation of the cosmic censorship conjecture for 
toroidal black holes}

\author{Marco M. Caldarelli\footnote{Electronic address: \sl
caldarel@science.unitn.it}} 

\address{Dipartimento di Fisica, Universit\`a di Trento, \protect\\ 
and Istituto Nazionale di Fisica Nucleare,\protect\\ 
Gruppo Collegato di Trento \protect\\ I-38050 Povo (TN), Italy}

\date{February 1998}

\maketitle

\begin{abstract}
{\small

We consider some aspects of quantum field theory of a conformally
coupled scalar field on the singular background obtained in the massless
limit of a class of toroidal black holes. 
The stress-tensor and its back reaction on the metric are computed using
point-splitting regularization, in the cases of transparent, Neumann
and Dirichlet boundary conditions. 
We find that the quantum fluctuations generate an event horizon
which hides the singularity. The resulting object can be interpreted as a
long lived remnant. We discuss the relevance of this
result in the context of the cosmic censorship conjecture, and 
in connection to the end point of the quantum evaporation process.}
\end{abstract}

\narrowtext

\section{Introduction}

Recently, interest has grown about topological black holes 
\cite{torusbh,ABHP,Mann,Vanzo,BLP}, namely solutions of
the vacuum Einstein equations with a negative cosmological constant, 
which are asymptotically anti--de~Sitter black holes, and have a compact event
horizon with a nontrivial topology. These black holes can result from
gravitational collapse \cite{MannSmith} and from quantum tunnelling processes
\cite{Mann1}. They can be charged \cite{cai1,BLP}, exist in dilaton gravity
\cite{cai1,cai}, and also admit a rotating generalization
\cite{KMV,LemosZanchin}. The reason why they violate the classical theorems
of general relativity forbidding nonspherical topologies is the presence
of the cosmological constant. 
For instance, the class of toroidal black holes is described by the metric
\begin{equation}
 {\rm d} s^2=-\left( -{2\eta\over r}+{r^2\over\ell^2}\right)\  {\rm d} t^2+
\left( -{2\eta\over r}+{r^2\over\ell^2}\right)^{-1} {\rm d} r^2+
r^2\left( |\tau|^2\  {\rm d} x^2+ {\rm d} y^2+2{\rm Re} \tau\  {\rm d} x
{\rm d} y\right)
\label{tbh}\end{equation}
where $x, y\in[0,1[$ and $\tau$ is the Teichm\"uller complex parameter
that determines the conformal class of the torus.
If $\eta>0$ the solution (\ref{tbh}) is a black hole of mass 
$M=\eta|{\rm Im}\ \tau|/4\pi$ with an
event horizon localized in $r_+=(2\eta\ell^2)^{1/3}$.
As the mass decreases to zero the event horizon shrinks and disappears for
$\eta=0$, leaving a naked singularity. We want to investigate this singular
background and check on its stability against the formation of an event
horizon when quantum fluctuations are taken into account in a 
semiclassical approach.
In particular we shall study the propagation of a conformally coupled scalar
field on this background and evaluate the back reaction on the metric, for
Dirichlet, Neumann and transparent boundary conditions, and show that
the quantum fluctuations of the field induce the formation of an event
horizon which hides the singularity.

The cosmic censorship conjecture states that, during
gravitational collapse, naked singularities never occur, at least for
asymptotically flat spacetimes \cite{Penrose,WaldGR}. 
However this conjecture is violated in presence of nontrivial topology;
Lemos has constructed a solution of collapsing matter with toroidal
symmetry that leads to the background (\ref{tbh}) with $\eta=0$, hence to a
naked singularity \cite{Lemos97}. It is interesting to investigate how
quantum mechanics modifies the issue of the collapse; we wonder if quantum
fluctuations, which cannot be neglected in the final stages of the
collapse, can prevent the classical naked singularities to form, hiding
them behind an event horizon.
If this shows to be the case, semiclassical back reaction of
quantum fields on the geometry would strengthen the cosmic censorship
conjecture. An instance of this is the case of extreme black holes, where
quantum emission drives the hole away from extremality \cite{Vanzo2}.
The issue of these questions is also fundamental in the problem of the
end point of evaporation. Black holes are known to evaporate by the
emission of thermal radiation \cite{Hawking}. 
In the case of toroidal topology, the black hole would
loose mass and end the evaporation process leaving a naked
singularity. Semiclassical results showing a dressing of the final
singularity would indicate that the final state could be a remnant, or perhaps
that the evaporation does not stop at $\eta=0$, but would continue through a
serie of dressed conical spacetimes, as has been suggested in three
dimensions \cite{LO}. 
However, in the final stages of evaporation, quantum effects are expected to
dominate, thus no final conclusion can be drawn in this sense without a
full theory of quantum gravity.

Analogous results have been obtained by various authors \cite{Steif,LO,SM,MZ}
for the non-rotating Ba\~nados-Teitelboim-Zanelli (BTZ) black hole
\cite{BTZ,BHTZ}.
This three-dimensional topological black hole shows the same behavior:
in the $M\rightarrow 0$ limit, the space-time acquires a naked singularity,
but quantum fluctuations produce a horizon at the semiclassical level.
We shall extend those results to the
four-dimensional case, which is physically more relevant. 
The analogy with the BTZ black hole is also formal: the
background (\ref{tbh}) with $\eta=0$ is the  quotient of the
four-dimensional anti--de~Sitter space with respect to a discrete group of
isometries. This fact simplifies the computation of the quantum stress-tensor
of a conformally coupled scalar field, and leads to interpret it as a
consequence of Casimir's effect.

\section{The geometry of the problem}

In this section we shall discuss the geometry of the background $ {\cal M}$ of
toroidal black holes, described by the singular metric
\begin{equation}
 {\rm d} s^2=-{r^2\over\ell^2}\  {\rm d} t^2 + {\ell^2\over r^2}\  {\rm d} r^2+
r^2\left(  {\rm d} X^2+ {\rm d} Y^2\right)
\label{ds}\end{equation}
where $t$ ranges over $ {\mathbb R}$, $r>0$ and $X, Y\in[0,1[$ with the
points $0$ and $1$ identified.
A first important remark about this metric
is that it is a locally anti--de~Sitter (AdS) metric and can be obtained
quotienting AdS space with respect to a discrete subgroup of its isometry
group. 

Let ${\cal V} =  {\mathbb R}^{3}_2$ be the flat 
pseudo-Euclidean space with coordinates $(x,y,z,u,v)$ endowed with the metric
$\eta_{ab}=(1,1,1,-1,-1)$ that defines the scalar product 
$(\cdot ,\cdot)_{\cal V}$. Anti--de~Sitter space is defined to be the
hypersuface \cite{HE}
\begin{equation}
AdS_4 = \left\{ {\mathbf x}\in{\cal V}\ | \ \eta_{ab}x^ax^a=-\ell^2\right\}
\end{equation}
endowed with the induced metric. To avoid closed timelike curves, we shall
work with the universal covering $\overline{\rm AdS}$ of anti--de~Sitter
space.
By construction, $\overline{\rm AdS}$ is a homogeneous and symmetric manifold,
with isometry group $SO(3,2)$. It solves Einstein equations with negative
cosmological constant $\Lambda=-{3/\ell^2}$ and has a constant negative
curvature $R=-{12/\ell^2}$.

Let us parametrize $\overline{\rm AdS}$ space by the coordinates $(t,r,X,Y)$
defined by 
\begin{equation}
t=-{\ell v\over u+x},\qquad
r=u+x,\qquad
X={y\over u+x},\qquad
Y={z\over u+x}.
\end{equation}
The metric assumes the form (\ref{ds}) in these coordinates, and is the
universal covering of the considered manifold. To obtain the background 
${\cal M}$ we need to make the identifications $X\sim X+m$ and $Y\sim Y+n$
with $m, n\in {\mathbb Z}$.
The manifold $ {\cal M}$ is regular everywhere except for $r=0$. This is not a
curvature singularity, but there are inextendible geodesics ending there as
the elements of the identification group leave this point fixed.
These identifications form an abelian subgroup $H\cong {\mathbb Z}\times
{\mathbb Z}\subset SO(3,2)$ of the homogeneity group. In the $ {\cal V}$
space the elements of the identification group are represented by the matrices
$$
\left(\Lambda^m\right)^a{}_b=\left(\begin{array}{ccccc}
1-{m^2\over 2} & -m & 0 & -{m^2\over 2} & 0\\ 
m & 1 & 0 &m&0\\ 
0&0&1&0&0\\ 
{m^2\over 2}&m&0&1+{m^2\over 2}&0\\ 
0&0&0&0&1
\end{array}\right)\qquad
\left(\Gamma^n\right)^a{}_b=\left(\begin{array}{ccccc}
1-{n^2\over 2} & 0 & -n & -{n^2\over 2} & 0\\ 
0 & 1 & 0 &0&0\\ 
n&0&1&n&0\\ 
{n^2\over 2}&0&n&1+{n^2\over 2}&0\\ 
0&0&0&0&1
\end{array}\right)
$$
corresponding respectively to the $X\mapsto X+m$ and the $Y\mapsto Y+n$
translations.
It can be proven that these matrices satisfy the abelian properties
$\Lambda^m\Lambda^n=\Lambda^{m+n}, \Gamma^m\Gamma^n=\Gamma^{m+n}$, and 
$\Lambda^m\Gamma^n=\Gamma^n\Lambda^m$,
and therefore we deduce that the manifold $ {\cal M}$ 
is the quotient $\overline{\rm AdS}/H$.

\section{Green functions}

Now we want to study the propagation of a conformally coupled scalar field 
$\phi$ on the background $ {\cal M}$ and calculate its Green function. 
The total action of the theory is $S=S_g+S_m$, where $S_g$ is the usual 
Hilbert-Einstein action for the gravitational field
\begin{equation}
S_g[g_ {\mu\nu}]={1\over 2\kappa}\int\sqrt{-g}\left( R-2\Lambda\right)
\end{equation}
with $\kappa=8\pi G$, and $S_m$ is the action of the matter field
\begin{equation}
S_m[g_ {\mu\nu},\phi]=- {1\over 2}\int\sqrt{-g}\left[g^{\mu\nu}
\nabla_\mu\phi\nabla_\nu\phi+\xi R\phi^2 \right]
\end{equation}
with $\xi=1/6$. The equations of motion are as usual
$G_ {\mu\nu}+\Lambda g_ {\mu\nu}=\kappa T_ {\mu\nu}$ and 
$\left(\square-\xi R\right)\phi=0$.

As $ {\cal M}$ is a quotient of $\overline{\rm AdS}$, 
its Green function can be obtained
from that of the covering space $\overline{\rm AdS}$ by the method of images. 
But $\overline{\rm AdS}$ is not a globally hyperbolic spacetime, therefore 
we have to impose boundary conditions at
infinity. To this end we conformally map $\overline{\rm AdS}$ to 
half of the Einstein
static universe (ESU). The boundary conditions at infinity become
conditions on the fields on the equator of ESU. As shown in \cite{AIS},
three choices of boundary conditions are possible; the transparent
boundary conditions are obtained quantizing the field using
smooth modes over the whole ESU, while the Neumann and Dirichlet boundary
conditions can be imposed on the equator of the ESU.
The Green functions are then
\begin{equation}
{\cal G}_{AdS}^\epsilon( {\mathbf x}, {\mathbf x}')
={1\over 4\pi^2}{1\over( {\mathbf x}- {\mathbf x}', 
{\mathbf x}- {\mathbf x}')_ {\cal V}}
+{\epsilon\over 4\pi^2}{1\over( {\mathbf x}+ {\mathbf x}', {\mathbf x}
+ {\mathbf x}')_ {\cal V}}
\label{G}\end{equation}
with $\epsilon = 0, +1, -1$ for transparent, Neumann and Dirichlet boundary
conditions respectively. The functional form of the Green functions is the
same as in Minkowski space because the background is conformally flat and the
field conformally coupled.
The second term in Eq. (\ref{G}) corresponds to an antipodal image in 
$\overline{\rm AdS}$ needed to force the boundary conditions.
The same results can be obtained by a direct mode sum \cite{BL,IO}.

Now we obtain the Green function with the appropriate boundary conditions
simply by summing over images, getting
\begin{equation}
{\cal G}_{ {\cal M}}^\epsilon( {\mathbf x}, {\mathbf x}')
={1\over 4\pi^2}\sum_{m,n\in {\mathbb Z}}\left[
{1\over( {\mathbf x}-\Lambda^m\Gamma^n {\mathbf x}', 
{\mathbf x}-\Lambda^m\Gamma^n {\mathbf x}')_ {\cal V}}
+{\epsilon\over( {\mathbf x}+\Lambda^m\Gamma^n {\mathbf x}', 
{\mathbf x}+\Lambda^m\Gamma^n {\mathbf x}')_ {\cal V}}\right].
\label{GM}\end{equation}
The series converges in the Dirichlet case, but diverges like $\sum[1/
(m^2+n^2)]$ for transparent and Neumann boundary conditions \cite{LO}. 
This is an infrared divergence, due to the existence of zero modes in
Neumann and transparent cases, that render the Klein-Gordon operator
non-invertible. Hence, we have to regularize the Green functions
eliminating these zero modes. This can be done defining the regularized
Green functions by 
\begin{equation}
{\cal G}_{ {\cal M}}^\epsilon( {\mathbf x}, {\mathbf x}'|s)
={1\over 4\pi^2}\sum_{m,n\in {\mathbb Z}}\left[
{1\over( {\mathbf x}-\Lambda^m\Gamma^n {\mathbf x}', {\mathbf x}
-\Lambda^m\Gamma^n {\mathbf x}')_ {\cal V}^s}
+{\epsilon\over( {\mathbf x}+\Lambda^m\Gamma^n {\mathbf x}', 
{\mathbf x}+\Lambda^m\Gamma^n {\mathbf x}')_ {\cal V}^s}\right]
\label{GMr}\end{equation}
which are convergent for $s>1$ and diverge in the $s\rightarrow 1$ limit.
Note that Eq. (\ref{GMr}) is a Green function in $(2s+2)$-dimensional
anti--de~Sitter space: we have used dimensional regularization.
In the calculation of the stress tensor we shall take derivatives of these
functions with respect to the spacetime variables, that will render finite the
series in $s=1$. In this point, the stress tensor will be the analytic
continuation of the dimensionally continued stress-tensor.

\section{Renormalized stress tensor}

The stress tensor for the field $\phi$ is given by
\begin{eqnarray}
T_ {\mu\nu}&\equiv&{-2\over\sqrt{-g}}\ {\delta S_m\over\delta g^ {\mu\nu}}
\nonumber\\
&=&{2\over 3}\ \nabla_\mu\phi\nabla_\nu\phi-{1\over3}\ 
\phi\nabla_\mu\nabla_\nu\phi
-{1\over 6}\ g_ {\mu\nu} g^{\alpha\beta}\nabla_\alpha\phi\nabla_\beta\phi
+{1\over 3}g_{\mu\nu}\phi\square\phi+{1\over 2\ell^2}\ g_{\mu\nu}\phi^2.
\label{TC}\end{eqnarray}
We observe first that the tensor $T_ {\mu\nu}$ is traceless, as is expected in
a classical conformally invariant theory. However, as we are in even
dimension, a conformal anomaly appears and breaks this
invariance at the quantum level.

Now, we aim to calculate the quantum stress tensor of the theory. We cannot
take the expectation value of formula (\ref{TC}) directly: after the
quantization procedure the field $\phi( {\mathbf x})$ becomes a 
distribution and expectation values like 
$\left<0\right|\phi( {\mathbf x})\phi( {\mathbf x})\left|0\right>$ 
are formally divergent quantities; 
we need to define a regularization procedure. We shall use the
point-splitting technique \cite{BirrellDavies}; in this framework, we
define the regularized stress tensor to be
\begin{eqnarray}
T_ {\mu\nu}( {\mathbf x})=\lim_{ {\mathbf x}'\rightarrow {\mathbf x}}
{1\over 3}&&\left[
\nabla_\mu\nabla_\nu'+\nabla'_\mu\nabla_\nu
-{1\over 2}\ \nabla_\mu\nabla_\nu- {1\over 2}\ 
\nabla_\mu'\nabla_\nu'\right.\nonumber\\&&
\left.- {1\over 2}\ g_ {\mu\nu}\left( g^{\alpha\beta}\nabla_\alpha\nabla_\beta'
+{1\over \ell^2}\right) 
\right]\left[ {\cal G}( {\mathbf x}, {\mathbf x}')- {\cal G}_{\rm sing}
( {\mathbf x}, {\mathbf x}')\right],
\end{eqnarray}
where $ {\cal G}_{\rm sing}( {\mathbf x}, {\mathbf x}')$ is a singular part
of the Green function, and we have used the equations of motion.

We have to distinguish two contributions to the Green function: the
former is the direct contribution, that is the Green function of 
$\overline{\rm AdS}$
space, the latter is the contribution coming from the sum over the images.
The $\overline{\rm AdS}$ contribution is divergent in the limit
$ {\mathbf x}'\rightarrow {\mathbf x}$ and we have to substract its
singular part to renormalize it. In fact, the stress tensor of the scalar
field in $\overline{\rm AdS}$ has to be proportional to the only tensor
available, i.e., the metric tensor $g_{\mu\nu}$. The proportionality
factor is one quarter of the conformal anomaly in our space, which is found
to be \cite{BirrellDavies} 
\begin{equation}
\langle\,T^\mu{}_\mu\rangle={\hbar\over 2880\pi^2}\left(
R_{ {\mu\nu}\alpha\beta}R^{ {\mu\nu}\alpha\beta}-R_{ {\mu\nu}}R^{ {\mu\nu}}
-\square R\right)=-{4\hbar\over 960\pi^2\ell^4},
\end{equation}
from which we deduce that the stress tensor contribution of
$\overline{\rm AdS}$ amounts to
\begin{equation}
\langle\,T_ {\mu\nu}^{AdS}\,\rangle=-{\hbar\over 960\pi^2\ell^4}\ g_ {\mu\nu}.
\end{equation}

In contrast, the contribution of the images remains finite after the
coincidence limit of the Green function. 
A calculation of the image contribution $\langle\,T^\pm_ {\mu\nu}\rangle$ 
in the  ${\cal V}$ space followed by a projection on $ {\cal M}$
\cite{Steif} yields 
\begin{equation}
\langle\,T^\pm_ {\mu\nu}( {\mathbf x})\rangle=\frac\hbar{6\pi^2}{\sum_{m,n
\in {\mathbb Z}}}'
\left( S^{\pm(m,n)}_ {\mu\nu}( {\mathbf x})-{1\over 4}\ g_ {\mu\nu}
g^{\alpha\beta}S^{\pm(m,n)}_{\alpha\beta}( {\mathbf x})\right),
\label{sti}\end{equation}
where by $T^\pm_ {\mu\nu}$ we mean respectively the stress tensor
calculated on the Green functions 
$(1/4\pi^2)( {\mathbf x}\pm {\mathbf x}', {\mathbf x}\pm {\mathbf
x}')_ {\cal V}^{-s}$, the prime on the sum
indicates that we drop the $(m,n)=(0,0)$ term and  $S^{\pm(m,n)}_ {\mu\nu}$ is
the pull-back
$S^{\pm(m,n)}_ {\mu\nu}=(\partial_\mu {\mathbf x}^a)(\partial_\nu 
{\mathbf x}^b)S^{\pm(m,n)}_{ab}$ of the
tensor in $T^*_ {\mathbf x} {\cal V}$ 
\begin{eqnarray}
S^{\pm(m,n)}_{ab}( {\mathbf x}|s)&=& 2s(s+1)\frac
{\left(\Lambda^m\Gamma^n\right)_a{}^cx_c
\left(\Lambda^{-m}\Gamma^{-n}\right)_b{}^dx_d
+\left(\Lambda^{-m}\Gamma^{-n}\right)_a{}^cx_c\left(\Lambda^m\Gamma^n\right)_b
{}^dx_d}{\left| x\pm\Lambda^m\Gamma^n x\right|^{2(s+2)}}\nonumber\\ 
&&-s(s+1)\frac{\left(\Lambda^m\Gamma^n\right)_a{}^cx_c\left(\Lambda^m\Gamma^n
\right)_b{}^dx_d
+\left(\Lambda^{-m}\Gamma^{-n}\right)_a{}^cx_c\left(\Lambda^{-m}\Gamma^{-n}
\right)_b{}^dx_d}
{\left| x\pm\Lambda^m\Gamma^n x\right|^{2(s+2)}}\nonumber\\ 
&&\mp s\frac{\left(\Lambda^m\Gamma^n\right)_{ab}+\left(\Lambda^m\Gamma^n
\right)_{ba}}
{\left| x\pm\Lambda^m\Gamma^n x\right|^{2(s+1)}}.
\end{eqnarray}
Now the series (\ref{sti}) is convergent in the point $s=1$, hence we can
safely remove the cutoff by setting $s=1$. Finally, the image contribution to
the stress tensor is given by Eq. (\ref{sti}), with
\begin{eqnarray}
S^{\pm(m,n)}_{ab}( {\mathbf x})&=&4\frac{\left(\Lambda^m\Gamma^n
\right)_a{}^cx_c
\left(\Lambda^{-m}\Gamma^{-n}\right)_b{}^dx_d
+\left(\Lambda^{-m}\Gamma^{-n}\right)_a{}^cx_c\left(\Lambda^m\Gamma^n
\right)_b{}^dx_d
}{\left| x\pm\Lambda^m\Gamma^n x\right|^6}\nonumber\\ 
&&-2\frac{\left(\Lambda^m\Gamma^n\right)_a{}^cx_c\left(\Lambda^m\Gamma^n
\right)_b{}^dx_d
+\left(\Lambda^{-m}\Gamma^{-n}\right)_a{}^cx_c\left(\Lambda^{-m}\Gamma^{-n}
\right)_b{}^dx_d}
{\left| x\pm\Lambda^m\Gamma^n x\right|^6}\nonumber\\ 
&&\mp\frac{\left(\Lambda^m\Gamma^n\right)_{ab}+\left(\Lambda^m\Gamma^n
\right)_{ba}}
{\left| x\pm\Lambda^m\Gamma^n x\right|^4}.
\end{eqnarray}
The total stress tensors with the various boundary conditions considered read
\begin{equation}
\langle\,T^\epsilon_ {\mu\nu}\rangle=\langle\,T^+_ {\mu\nu}\rangle+\epsilon
\langle\,T^-_ {\mu\nu}
\rangle+\langle\,T^{AdS}_ {\mu\nu}\,\rangle .
\end{equation}
Writing Eq. (\ref{sti}) in the coordinates $(t,r,X,Y)$ and performing the
pull-back we obtain
$$
\langle\,T^+_ {\mu\nu}\rangle=\frac{2\hbar r^4}{3\pi^2}{\sum_{m,n\in 
{\mathbb Z}}}'
{1\over (4\ell^2+(m^2+n^2)r^2)^3}
\ {\rm diag}\left(
(m^2+n^2)\ell^{-2};\ 0;\ 2m^2-n^2;\ 2n^2-m^2\right)
$$
and
$$
\langle\,T^-_ {\mu\nu}\rangle=\frac\hbar{2\pi^2 r^2}
{\sum_{m,n\in {\mathbb Z}}}'{1\over(m^2+n^2)^3}\ 
{\rm diag}\left(
-(m^2+n^2)\ell^{-2};\ (m^2+n^2){\ell^2\over r^4};\ n^2-3m^2;\ m^2-3n^2\right).
$$
From now on we define
\begin{equation}
\alpha\equiv{\sum_{m,n\in {\mathbb Z}}}'{1\over (m^2+n^2)^2}
={\sum_{m,n\in {\mathbb Z}}}'\frac{3m^2-n^2}{(m^2+n^2)^3}=6.0268\dots\,.
\end{equation}
Written in terms of the dimensionless variable $\zeta=r/2\ell$, the
stress-tensor reads
\begin{eqnarray}
&{\displaystyle\langle\,T^\epsilon_ {\mu\nu}\rangle=\frac\hbar{2\pi^2 \ell^2}
\ {\rm diag}}&{\displaystyle\left( 
-\frac\alpha{r^2}+\frac\epsilon{3\ell^2}\left( r\over 2\ell\right)^4E\left(r
\over 2\ell\right);\ 
{\alpha\ell^4\over r^6};\ 
-\frac{\alpha}{r^2}\ell^2+\frac\epsilon 6\left( r\over
2\ell\right)^4E\left(r\over 2\ell\right);\right.}\nonumber\\ 
&&{\displaystyle\left.
-\frac{\alpha}{r^2}\ell^2+\frac\epsilon 6\left( r\over
2\ell\right)^4E\left(r\over 2\ell\right)
\right)+\langle\,T_ {\mu\nu}^{AdS}\,\rangle}
\end{eqnarray}
where we have defined the auxiliary function (the derivative of an Epstein
function)
\begin{equation}
E(\zeta)\equiv\sum_{m,n\in {\mathbb Z}}{m^2+n^2\over\left(
1+(m^2+n^2)\zeta^2\right)^3}.
\end{equation}
The stress tensor we have obtained is regular and finite except for $r=0$
were it diverges. Furthermore, as can be easily verified, it is
conserved. Finally, we notice that the found stress tensor involves a
negative energy density, that may be interpreted it as a Casimir energy in
$ {\cal M}$ due to the identifications under the group $H$.

\section{Back reaction on the metric}

Now, we want to compute the semiclassical corrections to the geometry of
$ {\cal M}$ at order~$\hbar$ due to the presence of the quantum field
$\phi$. These corrections arise by solving the semiclassical equations
\begin{equation}
G_ {\mu\nu}+\Lambda g_ {\mu\nu}=\kappa\langle\,T_ {\mu\nu}\rangle.
\end{equation}

Let us first consider first the $\overline{\rm AdS}$ term of the stress
tensor. As $\langle\,T^{AdS}_ {\mu\nu}\rangle$ is proportional to the
metric, it only contributes to change the cosmological constant by a finite
renormalization: namely, 
\begin{equation}
G_ {\mu\nu}+\Lambda g_ {\mu\nu}=-{\hbar\kappa\over 960\pi^2\ell^4}\ g_ {\mu\nu}
\quad\Leftrightarrow\quad G_ {\mu\nu}+\Lambda_{ren} g_ {\mu\nu}=0
\end{equation}
where
\begin{equation}
\Lambda_{ren}=\Lambda+{\hbar\kappa\over 960\pi^2\ell^4}=-\frac
3{\ell^2}+{\hbar\kappa\over 960\pi^2\ell^4}.
\end{equation}
Having performed this renormalization we can forget the $\overline{\rm AdS}$
contribution in the stress tensor and consider only the contribution of the
images to the corrections to the metric. In the following we take $\ell$ to
be the renormalized constant.

We look now for a static solution, with toroidal symmetry, of the 
semiclassical equations. The general metric possessing these properties has
the form
\begin{equation}
 {\rm d} s^2 = - f^2(r)\  {\rm d} t^2+f^{-2}(r)\  {\rm d}
r^2+e^{2A(r)}\left( {\rm d} X^2+ {\rm d} Y^2\right).
\end{equation}
The semiclassical equations become
\begin{equation}
\left\{
\begin{array}{l}
\displaystyle -f^2\left[2f^2A''+(f^2)'A'+3f^2(A')^2+\Lambda\right]
=\kappa\langle\,T_{tt}\rangle\\ 
\\ 
\displaystyle f^{-2}\left[(f^2)'A'+f^2(A')^2+\Lambda\right]
=\kappa\langle\,T_{rr}\rangle\\ 
\\ 
\displaystyle 
e^{2A}\left[ {1\over 2}(f^2)''+(f^2)'A'+f^2A''+f^2(A')^2+\Lambda\right]=
\kappa\langle\,T_{XX}\rangle=\kappa\langle\,T_{YY}\rangle
\end{array}
\right.\label{sce}\end{equation}
As we want to study the corrections of order $\hbar$ to the
metric of $ {\cal M}$ we define
\begin{equation}
f^2(r)={r^2\over\ell^2}+\hbar\bar f(r)+{\cal O}(\hbar^2)\ ,\qquad
A(r)=\ln r+\hbar\bar A(r)+{\cal O}(\hbar^2)
\end{equation}
and substitute them in Eq. (\ref{sce}). The $0$-order terms satisfy the
equations
and cancel out, leaving, at first order, the linear equations
\begin{equation}
\left\{
\begin{array}{l}
\displaystyle -{\hbar r^2\over\ell^2}\left[{1\over r}\bar f'+{1\over r^2}\bar f
+2{r^2\over\ell^2}\bar A''+8\frac r{\ell^2}\bar A'\right]
=\kappa\langle\,T_{tt}\rangle\\ 
\\ 
\displaystyle {\hbar\ell^2\over r^2}\left[ {1\over r}\bar f'
+{1\over r^2}\bar f+4{r\over\ell^2}\bar A'
\right]=\kappa\langle\,T_{rr}\rangle\\ 
\\ 
\displaystyle 
\hbar r^2\left[ {1\over 2}\bar f''+{1\over r}\bar f'+{r^2\over \ell^2}\bar
A''+4{r\over \ell^2}\bar A'\right]=\kappa\langle\,T_{XX}\rangle
=\kappa\langle\,T_{YY}\rangle .
\end{array}
\right.\end{equation}
A linear combination of the first two equations yields
\begin{equation}
{ {\rm d}\over {\rm d} r}\left[ r^2\bar A'\right]=-{\kappa\ell^2\over 2\hbar}
\left[{\ell^2\over r^2}\langle\,T_{tt}\rangle+{r^2\over\ell^2}
\langle\,T_{rr}\rangle\right]
\end{equation}
which, integrated twice, gives
\begin{equation}
\bar A'(r)=-\frac{\epsilon\kappa}{12\pi^2(2\ell)^3} \left( 2\ell\over
r\right)^2E_1\left( r\over 2\ell\right)\ ,\qquad
\bar A(r)=-\frac{\epsilon\kappa}{12\pi^2(2\ell)^2} E_2\left( r\over
2\ell\right)
\end{equation}
where we have defined 
\begin{equation}
E_1(\zeta)\equiv\int \zeta^2E(\zeta)\, {\rm d} \zeta\ ,\qquad 
E_2(\zeta)\equiv\int {1\over \zeta^2}E_1(\zeta)\, {\rm d} \zeta .
\end{equation}
We do not need the exact form of $\bar A(r)$; but it can be proven that 
$\exp(2A)$ behaves asymptotically like $r^2$ and is regular everywhere. 
Substituting $\bar A$ in the Einstein equations we find the correction to
the lapse function
\begin{equation}
\bar f(r)=-{\alpha\kappa\over 2\pi^2 r^2}+ \frac {\epsilon\kappa}{3\pi^2\ell^2}
\left( 2\ell\over r\right) E_3\left( r \over 2\ell\right)
\end{equation}
with
\begin{equation}
E_3(\zeta)\equiv\int \zeta E_1(\zeta)\  {\rm d} \zeta.
\end{equation}
Performing the integrations and substracting an infinite integration constant
we find
\begin{eqnarray}
E_3(\zeta)={\sum_{m,n\in {\mathbb Z}}}'{}&&\left[
\left({\zeta^2\over 16\left( m^2+n^2\right)^{1/2}}
-{3\over 16\left( m^2+n^2\right)^{3/2}}\right)
\left(\arctan\left(\sqrt{m^2+n^2}\zeta\right)-
{\pi\over 2}\right)+\right.\nonumber\\ 
&&\,\,\,\left.+{\zeta\over 16(m^2+n^2)}
+{\zeta\over 8(m^2+n^2)
\left( 1+(m^2+n^2)\zeta^2\right)}\right]+\eta.
\end{eqnarray}
The asymptotic behavior of this function is
\begin{equation}
E_3(\zeta)={\alpha\over 3\zeta}+\eta+{\cal O}\left({1\over\zeta^3}\right),
\end{equation}
from which we deduce the asymptotic form of the lapse function 
\begin{equation}
\left\{\begin{array}{ll}
\displaystyle 
f^2(r)={r^2\over \ell^2}-{1\over 18}
{\hbar\kappa\alpha\over\pi^2r^2}
+\frac{\hbar\eta} r +{\cal O}\left({\hbar\over r^4}\right)&\quad (N)\\ 
\\ 
\displaystyle
f^2(r)={r^2\over \ell^2}-{17\over18}
{\hbar\kappa\alpha\over\pi^2r^2}
+\frac{\hbar\eta} r +{\cal O}\left({\hbar\over r^4}\right)&\quad (D)
\end{array}\right.\label{lf}
\end{equation}
for Neumann and Dirichlet boundary conditions respectively.
Hence, the integration constant $\eta$ corresponds to the physical mass of
the black hole. It is not fixed by the semiclassical theory, but we can
set it equal to zero, to match asymptotically the corrected metric with its
classical background, and to obtain a massless solution.

Thus the semiclassical metric acquires a curvature singularity in the 
origin and the perturbation due to the propagation of the quantum field
$\phi$ develops a horizon that hides the singularity.

This is easily seen in the case of transparent boundary conditions
($\epsilon=0$), where a horizon forms in
$r_+=\left( \kappa\alpha\hbar\ell^2\right)^{1/4}$. In terms of the
Planck length $\ell_p$ the black hole radius goes like $\sqrt{\ell\ell_p}$
and formally has a Hawking temperature \cite{Hawking}
\begin{equation}
T={\hbar r_+\over\pi\ell^2},
\end{equation}
which is immensely smaller than the temperature of the initially
macroscopic toroidal black hole. If the dressed singularity continues to
evaporate, then it is a kind of long lived remnant.
When Neumann or Dirichlet boundary conditions are imposed, the expression 
of the lapse function is more complicated, and we cannot find
analytically the location of the horizon. However, it is still possible to
prove that in the perturbed space-time the singularity is hidden behind an
event horizon.
We start from the fact that $E_3(\zeta)$ is decreasing for positive $\zeta$
and assumes in 0 the value
\begin{equation}
E_3(0)={3\pi\over 32}\alpha_{3/2} = 2.66\ldots\qquad
\alpha_{3/2}\equiv{\sum_{m,n\in {\mathbb Z}}}'\left( m^2+n^2\right)^{-\frac
32}=9.033\ldots\,. \end{equation}
Thus, we obtain the inequality
\begin{equation}
0<E_3(\zeta)\leq{3\pi\over 32}\alpha_{3/2}\ ,\qquad\zeta\geq 0.
\end{equation}
Substituting it in the expression (\ref{lf}), we find that the  lapse
function (for Neumann and Dirichlet boundary conditions respectively)
is resticted to vary between two simple functions
\begin{equation}
\left\{
\begin{array}{ll}
\displaystyle {r^2\over\ell^2}-{\hbar\kappa\alpha\over 2\pi^2r^2}\leq f^2(r)
\leq {r^2\over\ell^2}-{\hbar\kappa\alpha\over 2\pi^2r^2}
+{\hbar\kappa\alpha_{3/2}\over 16\pi\ell r}&\quad(N)\\ 
\\ 
\displaystyle {r^2\over\ell^2}-{\hbar\kappa\alpha\over 2\pi^2r^2}
-{\hbar\kappa\alpha_{3/2}\over 16\pi\ell r}\leq f^2(r)
\leq {r^2\over\ell^2}-{\hbar\kappa\alpha\over 2\pi^2r^2}&\quad(D)
\end{array}
\right.
\label{disug}\end{equation}
that are negative for sufficiently small $r$ and become positive as $r$ grows.
This is more evident if we multiply Eq. (\ref{disug}) by $\ell^2r^2$:
\begin{equation}
\left\{
\begin{array}{ll}
\displaystyle r^4-{\hbar\kappa\alpha\ell^2\over 2\pi^2}\leq \ell^2r^2f^2(r)
\leq r^4-{\hbar\kappa\alpha\ell^2\over 2\pi^2}
+{\hbar\kappa\alpha_{3/2}\ell\over 16\pi}r&\quad(N)\\ 
\\ 
\displaystyle r^4-{\hbar\kappa\alpha\ell^2\over 2\pi^2}
-{\hbar\kappa\alpha_{3/2}\ell\over 16\pi}r \leq \ell^2r^2f^2(r)
\leq r^4-{\hbar\kappa\alpha\ell^2\over 2\pi^2}&\quad(D).
\end{array}
\right.
\label{dis}\end{equation}
Since a quartic equation of the form $x^4+ax-b=0$, with $b>0$,
has a unique real and positive root, inspection into Eqs. (\ref{dis})
shows that the lapse function is zero in $r_+^N$ ($r_+^D$) for Neumann
(Dirichlet) boundary conditions, negative for lower values of $r$ and
positive for higher values.
Furthermore, the zeroes of the lapse function satisfy $r^D_+>r_+>r^N_+$, 
and, as $\ell$ grows, $r_+^D$ and $r_+^N$ approach $r_+$.
Thus, the quantum fluctuations dress the naked singularity. For a Neumann
field we find the horizon nearer to the singularity, while Dirichlet boundary
conditions push the horizon outwards, and transparent boundary conditions
give rise to an horizon located between the other two. 

\section{Conclusions}

In this paper we have analyzed the propagation of a conformally coupled
scalar field on the singular manifold ${\cal M}$. The main result is that the
quantum fluctuations dress the singularity. We have argued that this
quantum dressing could be a general feature of propagation of quantum
fields on singular backgrounds generated by collapsing matter, and would 
imply that quantum effects would work in the sense of the cosmic censorship
conjecture.
Furthermore, as $ {\cal M}$ appears to be the natural end point of the
evaporation process, the semiclassical insight seems to discredit the
possibility of naked singularities as end point of evaporation. However, as
the horizon enjoys thermal properties, the best we can conclude is that the
dressed singularity behaves as a kind of long lived remnant.

A further extension of these results could be the inclusion of the
Teichm\"uller complex parameter, which specifies conformally equivalent
classes of the torus, and see if it introduces significant changes; another
interesting question could be the generalization of the present discussion to
other kinds of field, like conformal Weyl spinors fields or vector fields.

\section*{acknowledgments}

I would like to thank Valter~Moretti, Luciano~Vanzo, and Sergio~Zerbini for
many valuable discussions.

\end{document}